# On the variational principle for fractional kinetic theory

Sumiyoshi Abe[1,2] and Akifumi Oohata[1]

[1] Department of Physical Engineering, Mie University, 514-8507, Japan

[2] Institute of Physics, Kazan Federal University, Kazan 420008, Russia

**Abstract.** In a recent paper (Abe S 2013 *Phys. Rev.* E **88** 022142), a variational principle has been formulated for spatiotemporally-fractional Fokker-Planck equations and applied to derivations of their approximate analytic solutions based on the Lévy Ansatz. Here, the problem of the constraint associated with normalization condition on a probability distribution behind the principle is discussed. It is shown that the action functional possesses a specific transformation property in terms of an auxiliary field and the constraint turns out to have already been imposed implicitly in terms of such a structure.

# 1. Introduction

Variational principle has been playing an outstanding role in science and will be doing so over the future. It offers a unified approach to deriving basic equations as stationarity conditions on relevant functionals. In physics, functionals are often actions that are given in terms of Lagrangians or Hamiltonians. Therefore, time-reversal invariance is essential, leading to an obvious problem for systems without such an invariance, e.g., dissipative systems. This point can, however, be overcome if the space of variables is extended.

In a recent work [1], a variational principle has been formulated for spatiotemporally-fractional Fokker-Planck equations in the context of anomalous diffusion and applied to systems of physical interest as a method for obtaining approximate analytic solutions with the help of the Lévy Ansatz. There, an auxiliary field has been introduced in order to extend the space of variables because of the absence of time-reversal invariance in the equations. However, an issue regarding the constraint associated with the normalization condition on a probability distribution has not been discussed in that work. The purpose of this short note is to clarify this point in connection with a specific transformation property of the action.

# 2. Fractional Fokker-Planck equation

The fractional Fokker-Planck equation discussed in Ref. [1] is of the following form:

$$\frac{\partial p(x,t)}{\partial t} = {}_0D_t^{1-\alpha}\left\{-\frac{\partial}{\partial x}[F(x)p(x,t)] - D^*(-\Delta)^{\gamma/2}p(x,t)\right\}. \tag{1}$$

Here, $p(x,t)\,dx$ is the probability of finding a particle (a random walker) in the interval $[x, x+dx]$ at time $t$. The domain of $p(x,t)$ is $(-\infty,\infty)\times[0,T]$. $D^*$ is a generalized diffusion coefficient, and the first term inside the braces on the right-hand side is a drift. ${}_0D_t^{1-\alpha}$ and $-(-\Delta)^{\gamma/2}$ are the Riemann-Liouville fractional differential operator and the Riesz fractional "Laplacian" [2,3], respectively. The ranges of $\alpha$ and $\gamma$ of physical interest are

$$0 < \alpha < 1, \qquad 0 < \gamma < 2. \tag{2}$$

Then, the Riemann-Liouville operators are defined by

$$_0D_t^{1-\alpha} f(t) = \frac{1}{\Gamma(\alpha)}\frac{d}{dt}\int_0^t ds\,(t-s)^{\alpha-1} f(s), \tag{3}$$

$$_tD_T^{1-\alpha} f(t) = \frac{1}{\Gamma(\alpha)}\left(-\frac{d}{dt}\right)\int_t^T ds\,(s-t)^{\alpha-1} f(s), \tag{4}$$

which fulfill the following relation:

$$\int_0^T dt\left[{}_0D_t^{1-\alpha}f(t)\right]g(t) = \int_0^T dt\, f(t)\left[{}_tD_T^{1-\alpha}g(t)\right]. \tag{5}$$

On the other hand, the Riesz operator is defined by

$$-(-\Delta)^{\gamma/2} = -\frac{1}{2\cos(\pi\gamma/2)}\left[\frac{d^{\gamma}}{dx^{\gamma}} + \frac{d^{\gamma}}{d(-x)^{\gamma}}\right], \tag{6}$$

which satisfies

$$\int_{-\infty}^{\infty} dx \left[-(-\Delta)^{\gamma/2}\psi(x)\right]\phi(x) = \int_{-\infty}^{\infty} dx\, \psi(x)\left[-(-\Delta)^{\gamma/2}\phi(x)\right]. \tag{7}$$

It is noted that the Riesz differentiation of a constant vanishes, whereas the Riemann-Liouville differentiation does not.

The physical mechanism underlying Eq. (1) is described in the language of continuous-time random walks [4] as follows. Temporal fractionality characterized by $\alpha$ implies that the waiting-time distribution decays as a power law, $P_W(t) \sim t^{-1-\alpha}$, and does the jump distribution too, $P_J(x) \sim |x|^{-1-\gamma}$. The power-law waiting (jumping) suppresses (enhances) diffusion. Therefore, resulting the diffusion property is characterized by the interplay between these two effects. In particular, presence of long jumps has motivated the work in Ref. [1] to propose the Lévy Ansatz. There, a periodic drift is considered as an example and, with the help of the Rayleigh-Ritz-like method, it is shown that motion of the center of the Lévy distribution is slow (i.e., not an exponential-law behavior familiar in ordinary kinetic theory but a power-law one) and the diffusion property is characterized by $l \sim t^{\alpha/\gamma}$, where $l$ stands for the spatial extension of the distribution such as the half width. A particular case with $\alpha/\gamma = 1/2$ is a realization of the so-called non-Gaussian normal diffusion.

## 3. Action and variational principle

Clearly, the fractional Fokker-Planck equation (1) is not invariant under time reversal. To construct the corresponding action functional, an auxiliary field $\Lambda(x,t)$ has been introduced in Ref. [1] in order to extend the space of variables. It is noted that $\Lambda(x,t)$ is not a probability distribution. The action presented reads

$$I[p,\Lambda] = -\int_0^T dt \left\langle \dot{\Lambda} + {}_t D_T^{1-\alpha} \left\{ F\Lambda' - D^*(-\Delta)^{\gamma/2}\Lambda \right\} \right\rangle - \left\langle \Lambda \right\rangle\big|_{t=0}, \qquad (8)$$

where $\dot{\Lambda} = \partial \Lambda / \partial t$, $\Lambda' = \partial \Lambda / \partial x$ and $\langle A \rangle = \int_{-\infty}^{\infty} dx\, A(x,t) p(x,t)$. Introduction of the last term on the right-hand side is inspired by the works in Refs. [5,6], where a variational principle for the Liouville-von Neumann equation is discussed (note that the Liouville-von Neumann equation possesses time-reversal invariance in contrast to the present fractional Fokker-Planck equation).

The variations of this action with respect to $\Lambda$ and $p$ are found to be given as follows:

$$\delta_\Lambda I = \int_0^T dt \int_{-\infty}^{\infty} dx \left\{ \dot{p} + {}_0 D_t^{1-\alpha}\left[ (Fp)' + D^*(-\Delta)^{\gamma/2} p \right] \right\} \delta\Lambda - \int_{-\infty}^{\infty} dx\, p\,\delta\Lambda \bigg|_{t=T}, \qquad (9)$$

$$\delta_p I = -\int_0^T dt \int_{-\infty}^{\infty} dx \left\{ \dot{\Lambda} + {}_t D_T^{1-\alpha}\left[ F\Lambda' - D^*(-\Delta)^{\gamma/2}\Lambda \right] \right\} \delta p - \int_{-\infty}^{\infty} dx\, \Lambda\,\delta p \bigg|_{t=0}, \qquad (10)$$

where the relations in Eqs. (5) and (7) have been used. The temporal boundary terms in these equations vanish if $\delta\Lambda(x,T) = 0$ and $\delta p(x,0) = 0$. To realize them, let us impose the following temporal boundary conditions:

$$\Lambda(x, T) = \Lambda_0(x), \tag{11}$$

$$p(x, 0) = \delta(x - X_0), \tag{12}$$

where $\Lambda_0(x)$ is a fixed function and $X_0$ is a constant. Actually, $p(x,0)$ in Eq. (12) can be replaced by an arbitrary normalized distribution, but this point does not affect the subsequent discussion.

In Ref. [1], it is claimed that the stationarity conditions, $\delta_\Lambda I = 0$ and $\delta_p I = 0$, lead to Eq. (1) and

$$-\frac{\partial \Lambda(x,t)}{\partial t} = {}_t D_T^{1-\alpha} \left\{ F(x) \frac{\partial \Lambda(x,t)}{\partial x} - D^* (-\Delta)^{\gamma/2} \Lambda(x,t) \right\}, \tag{13}$$

respectively.

It is of interest to observe that the field equation (13) has a structure similar to Eq. (1) with time reversal. In particular, the fractional operator ${}_t D_T^{1-\alpha}$, which is the adjoint of ${}_0 D_t^{1-\alpha}$ in Eq. (1), appears through the relation in Eq. (5). This can be thought of as the invariance in the present system.

4. Transformation property of the action and normalization constraint

Now, the problem in the above scheme is that the variation with respect to the probability distribution should be performed under the constraint associated with the normalization condition

$$\int_{-\infty}^{\infty} dx\, p(x,t) - 1 = 0, \tag{14}$$

which is preserved in time by Eq. (1). Below, we shall see that this problem is indivisibly related to a specific transformation property that the action possesses.

A key point is in the fact that the auxiliary field $\Lambda$ is not a probability distribution and does not have to be normalized. [Even if $\Lambda$ is forced to be normalized at certain time, the field equation (13) does not preserve such a condition.] This allows us to transform the field. The transformation we consider here is the following one:

$$\Lambda(x,t) \to \Lambda(x,t) + \int_t^T ds\, \lambda(s). \tag{15}$$

It is important to note that this keeps the final condition in Eq. (11) unchanged. Under this, the action in Eq. (8) is found to transform as follows:

$$I[p, \Lambda] \to I[p, \Lambda] + \int_0^T dt\, \lambda(t) \left[ \int_{-\infty}^{\infty} dx\, p(x,t) - 1 \right], \tag{16}$$

where Eq. (12) has been used. Identifying $\lambda$ with a Lagrange multiplier, we see that the second term is precisely the constraint associated with the normalization condition in Eq. (14). In other words, even if the constraint would be included in the action, it could be eliminated by the inverse transformation of Eq. (15). Therefore, the constraint turns out to have already been imposed implicitly in the action (8) through additive arbitrariness of $\Lambda$, which is somewhat analogous to the gauge-theoretic structure. This justifies the scheme in Section 3.

**5. Concluding remarks**

We have reexamined the variational principle that has recently been developed for spatiotemporally-fractional Fokker-Planck equations. In particular, we have discussed the problem of the constraint associated with the normalization condition behind the principle. We have shown that the action possesses a specific transformation property concerning additive arbitrariness of the auxiliary field and the constraint has already been imposed implicitly.

In Ref. [6], it is pointed out for the variational principle for the Liouville-von Neumann equation that the constraint of the normalization condition on a density matrix can be taken care of if the phase of an auxiliary matrix is redefined. This idea may be analogous to ours expressed in Eq. (15).

Besides its useful role in obtaining approximate analytic solutions of kinetic equations, a variational principle also provides a powerful theoretical framework for discussing symmetry of probability distributions. It is shown e.g. in Refs. [7,8] how power-law distributions as the solutions of a class of non-fractional Fokker-Planck equations can be related to dilatation symmetry [note that in those papers, the temporal boundary terms like in Eq. (8) are not contained]. It is of interest to develop such a symmetry discussion for fractional kinetics.

**Acknowledgments**

One of the authors (S. A.) would like to thank the organizers of The 4th International Workshop on Statistical Physics and Mathematics for Complex Systems (12-16 October, 2014, Yichang, China) for inviting him to give a lecture. His work has been supported in part by a Grant-in-Aid for Scientific Research from the Japan Society for the Promotion of Science and by the Ministry of Education and Science of the Russian Federation (federal program of competitive growth of Kazan Federal Univesity).